\documentclass[prd,superscriptaddress,onecolumn,showpacs,11pt,msmath,preprintnumbers,showkeys]{revtex4}
  \usepackage{wrapfig,rotating}
  \usepackage{graphicx,epsfig,color}

  \makeatletter
  
  \newenvironment{figurehere}
  {\def\@captype{figure}}
  {}
  \makeatother

  \usepackage{graphicx}% Include figure files
  \usepackage{dcolumn}% Align table columns on decimal point
  \usepackage{bm}% bold math

  \def\beq{\begin{equation}}
  \def\eeq{\end{equation}}
  \def\beeq{\begin{eqnarray}}
  \def\eeeq{\end{eqnarray}}

  \def\2GPD{$_2\mbox{GPD}$}
    \def\nn{$1\otimes 2$}
  \def\12{$1\otimes 2^u$}
  \def\ss{$2\otimes 2$}
  \def\21{$1\otimes 2^d$}

  \def\Qsep{Q_{\mbox{\rm\scriptsize sep}}}
  \def\Qsep2{Q^2_{\mbox{\rm\scriptsize sep}}}
  
\begin{document}

  \title{Direct versus resolved photons in DPS in photoproduction on proton and nuclei.}

 \author{B.\ Blok ,
 R. Segev \\ \normalsize  Department of Physics, \\Technion -- Israel Institute of Technology,\\
 Haifa, Israel}
 %\abstract
 \begin{abstract}
We study the process of double parton scattering (DPS)  associated with  the photoproduction at  a future  Electron-Ion collider (EIC)  and HERA.  We  show that in the case of 
the resolved photon   the \nn $\,\,$
 processes  lead, even at small transverse momenta of the hard processes ,    to the increase of the DPS by a factor of order 1.6  in the  significant  part of the phase space,
relative to the predictions of the mean field based models . Moreover we 
study the kinematic region where direct photon contribution is dominant  and show it s  boundaries for charm and light quark jets.
For charmed jets we see that the relevant region is $x_\gamma\ge 0.2-0.4$. This region is even enhanced if we consider   the photoproduction on the nuclei.

 \end{abstract}
   \maketitle
 \thispagestyle{empty}

 \vfill

\section{Introduction}
\par The first theoretical works on double parton scattering (DPS) relate to early eighties  \cite{TreleaniPaver82,mufti}. At the same time the DPS  was first observed experimentally 
at  Tevatron. Significant improvement in our knowledge of the DPS occurred in the Large Hadron Collider (LHC)  era.  In particular, from the theoretical  point of view,
 the systematic pQCD formalism to study DPS was developed.\cite{stirling,BDFS1,Diehl,stirling1,BDFS2,Diehl2,BDFS3,BDFS4,Diehl:2017kgu,Manohar:2012jr,ST,BSW} 
, see  the recent review \cite{book}.
\par The recent applications however dealt mostly with pp and pA collisions in the central and close to the central kinematics, as they are observed at LHC.
On the other hand,  the multi-parton interactions (MPI), including DPS , in deep-inelastic scattering (DIS) and photoproduction attracted relatively little attention, 
except some research
on MPI in HERA, where it was claimed that the inclusion of MPI improves the agreement between experiment and theoretical 
description of HERA measurements \cite{Yung,jimmy1}, although no concrete evidence for MPI in HERA was presented experimentally
\cite{Yung1,Yung2}.
\par The situation however started to change recently due to the expected  start  of the  Electron-Ion collider (EIC) in Brookhaven and further progress in experimental and theoretical studies of 
the ultraperipheral
collisions at LHC \cite{UPC}.
More recently,
the multi-parton interactions in DIS  and photoproduction for resolved photons were studied in \cite{CR} and in \cite{jimmy2}, where the contribution of MPI in the underlying event 
in the  future Electron-Ion collider  was considered.
\par 
 More recently the 
DPS  in   the DIS and photoproduction for  direct  photons was  studied in  \cite{BS,BlokSegev}. In particular, it was found  that a large number of DPS events can be observed both in photoproduction and electroproduction in the new 
EIC collider due to its large luminosity.
\par The  new feature of the pQCD approach to DPS, relative to parton model,  is the existence , in addition to conventional \ss $\,\,$ mean field mechanism   for DPS  (Fig. \ref{dps1}left) ,  of the  \nn$\,\,$  mechanism, Fig. \ref{dps1} center and right,  see Ref. \cite{BDFS2}.  In  the \nn  $\,\,$mechanism   
the parton from the proton\ resolved photon  wave function splits perturbatively into the two partons,  each  of them initiates  a new hard process.
\begin{center}
\begin{figurehere}
 \includegraphics[height=4.5cm,angle=0]{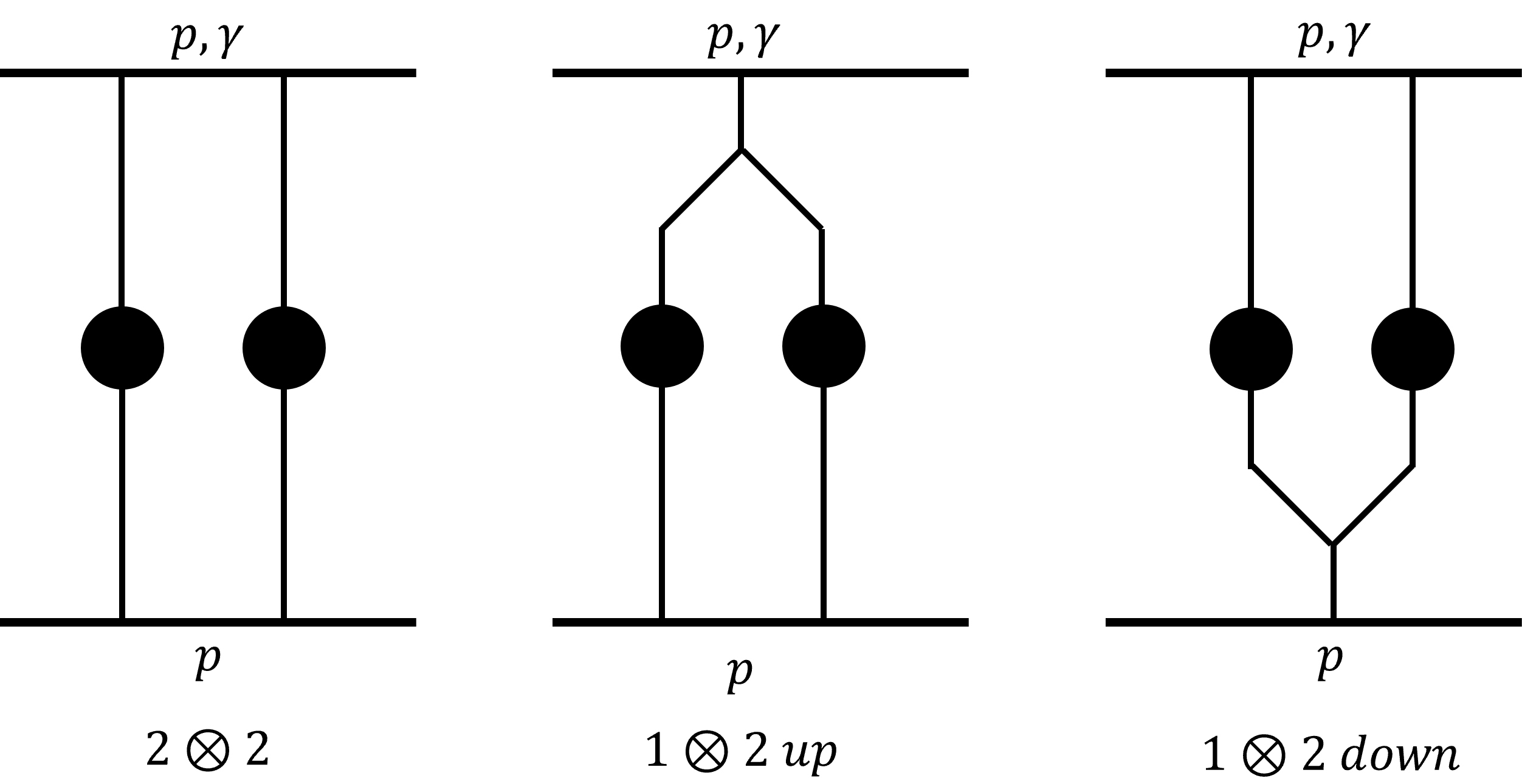}
\caption{\label{dps1} The  \ss, (left) \12 (center) ,\21 (right) processes in  DPS, the two initiated hard processes are depicted by two black blobs.}
\end{figurehere}
\end{center}

The $\gamma p$  collisions with the direct 
photon are the way to explicitly see the \nn   mechanism, see Fig. \ref{dps2}.
\begin{center}
\begin{figurehere}
 \includegraphics[height=4.5cm,angle=0]{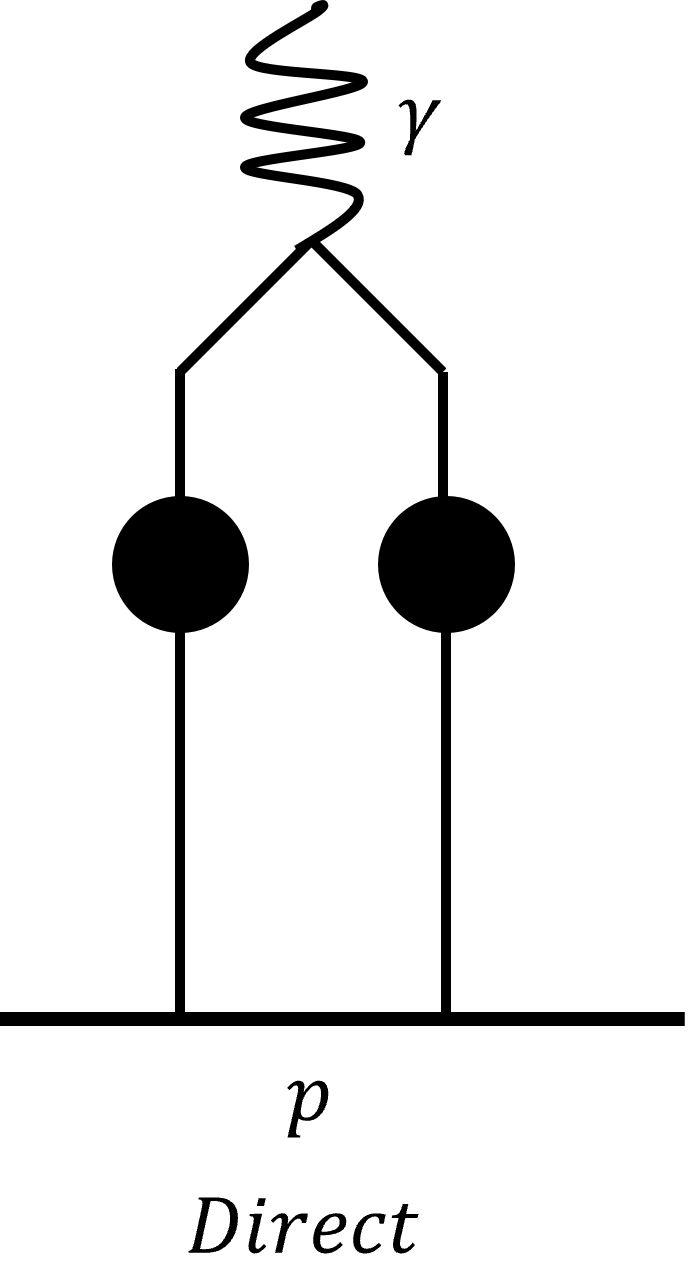}
\caption{\label{dps2} The DPS for direct photon.}
\end{figurehere}
\end{center}
\par The DPS events  occur for both the direct and resolved photons. Here by the resolved photon we understand the photon with multicomponent  QCD wave function. The properties of such object 
can be studied using vector dominance \cite{Feynman}. The DPS theory of the resolved photon is analogous to the DPS theory  of pp collisions, the main difference being the lack of symmetry between target and projectile.
Consequently there are conventional \ss $\,\,$mechanism, and  \nn $\,\,$mechanism. Due to the lack of symmetry we denote the process when the parton from the wave function of the reeolved photon 
splits into two hard partons as \12 process, and when the parton from the wave function of the target nucleon spits into two we denote this process as  \21 $\,\,$ process, see Fig. \ref{dps1}.
\par On the other hand by direct photon we understand the process when the photon directly transforms  into quark-antiquark dipole, that initiates two hard processes,
each in the collision of a hard quark and a hard parton from the target nucleon. The corresponding cross sections were studied for both photo and electroproduction in \cite{BS,BlokSegev}.

\par The purpose of this paper is twofold. First we consider the relative contribution to DPS of direct and resolved photons for the same final state.. 
We show that  for the final state of the hard processes, each consisting of a  charmed jets and of a  gluonic jet, considered in \cite{BS,BlokSegev} ,
the direct photons become dominant for $x_{\gamma}\equiv x_1+x_2\ge 0.3-0.4$ ( and even for smaller $x_\gamma$ for $\gamma A$ collisions with the same energy and luminosity).
For smaller $x_\gamma$ the direct photon contribution  rapidly decreases, and  the main contribution to DPS comes from resolved photons. Here $x_1,x_2$ are the Bjorken fractions of the jets
in the hard process: $Q_1=x_1q+x_3p+Q_{t1}, Q_2=x_2p+x_3q+Q_{t2} $.  where q is the photon and p is the nucleon momentum, $Q_1,Q_2$ are the hard jet momenta in these processes.
In our calculations we assume $p^2=q^2=0$, i.e. we neglect the proton mass.
\par On the other hand we see that for the case of the final state with two  hard processes , each including light quark jet and  gluon jet, the contribution of resolved and direct photons becomes of the same order only for large $x_\gamma\sim 0.9$. Nevertheless
it is significant for all $x_\gamma\ge 0.3-0.4$, and thus must be included in Monte Carlo generators like JIMMY and its updated versions \cite{jimmy2}.
\par Our results indicate that at least in some kinematic region for two charm-- gluon jet  final states we indeed can separate the direct photons.
\par Second, we study the resolved photons, in particular we consider the same final states as for direct photon and in addition the final state that includes two gluon dijets. We see that the corresponding 
DPS cross sections reach maximum for $x_\gamma \sim 0.01-0.1$, depending on energy, and decrease for smaller $x_\gamma$ due to phase constraints.
\par We for the first time directly calculate \12 $\,\,$ and \21  $\,\,$ processes   contributions  to the DPS cross sections, We see that the \ss $\,\,$ mechanism is dominant (due to small available transverse momenta).
. However for wide kinematic region the \12$\,\,$ and \21 $\,\,$ processes  increase the DPS cross section by a factor of up to 1.6 and thus must be taken into account in any DPS simulation.
We expect from the experience of pp collisions, that the role of these mechanisms  will further increase with the increase of transverse momenta.
\par The paper is organised in the following way. In section 2 we  review the basic formalism developed in \cite{BDFS1,BDFS2,BDFS3,BlokSegev}  to calculate the DPS in the case
of the direct and resolved photons. In section 3 we study \12$\,\,$ and \21$\,\,$  contributions to resolved photon DPS numerically. In section 4 we compare numerically direct and resolved photon contributions to the  total DPS cross sections.
In section 5 we consider the case of the nuclei, and our results are summarized in conclusion.

\section{Basic Formalism.}
\par In this section we shall briefly review the basic formalism developed in \cite{BDFS1,BDFS2,BDFS3,BS,BlokSegev} to calculate DPS both for resolved and 
for direct photon.
\par The cross section for DPS photoproduction is  proportional to photon flux, which is given, for the case of the photoproduction by  \cite{Frixione:1993yw}
\beq
n_\gamma (y)=\frac{\alpha_{\rm em}}{2\pi}y(\frac{(1 + (1 - y)^2)}{y^2}\log[\frac{Q_{\rm max}^2}{Q_{\rm min}^2(y, m)}]
   -2m_e^2(1/Q_{\rm max} ^2- 1/Q_{\rm min}(y, m])^2)
\eeq
Here  $y $ is a part of the energy (squared)  of the electron proton system $W^2$   that goes to the photon--proton system:
\beq
s=yW^2
\eeq
\beq 
Q_{\min}^2=m_e^2y^2/(1-y)
\eeq
and $Q_{\rm max}\sim 1$ GeV. Here $m_e=0.5 $  MeV is the mass of the electron, $\alpha_{\rm em}=1/137$ is the electromagnetic coupling.
The photon can be either resolved, or direct. 

\subsection{Resolved  Photon.}
\subsubsection{\ss$\,\,$ calculation.}
There are two contributions (see Fig. \ref{dps1}) ,  usually called  \ss $\,\,$ and   $1\otimes 2$ $\,\,$ mechanisms. 
\par Let us start from \ss $\,\,$ contribution for the resolved photon.  Recall first the formalism for the pp collisions.The contribution of the \ss $\,\,$ mechanism  to the cross section 
of DPS  is given 
by \cite{BDFS1}
\beq
\sigma=\int \frac{d^2\Delta}{(2\pi)^2} G_2(x_1,x_2,Q^2_1,Q^2_2,\Delta )_2G_2(x_3,x_4,Q^2_1,Q^2_2,\Delta )\frac{d\sigma_{13}}{dt_1}\frac{d\sigma_{24} }{dt_2} \label{gpd1}  \eeq
where $G_2$ are the  generalized  double  parton distribution functions  ($GPD_2$)  of the  proton. The $Q_1,x_1,x_3$ 
and the  $Q_2,x_2,x_4$  are the transverse momenta and Bjorken fractions  of the two hard processes that contribute to DPS, $d\sigma/dt_{1,2},t_1=Q_1^2, t_2=Q_2^2$ are the  differential cross sections
of the corresponding hard processes (see i.e. \cite{Webber}. for explicit formulae 
 for  the differential cross sections of hard dijet processes) . The momentum $\vec \Delta$ is the two dimensional momenta conjugated to the transverse distance between two partons.
 
\par In the mean field approximation the proton GPD$_2$ can be written as \cite{BDFS1}
\beq
G_2(x_1,x_2,Q^2_1,Q_2^2, \Delta)=G_1(x_1,Q_1^2,\Delta)G_1(x_2,Q_2^2,\Delta)
\eeq
where $G_1$ are the Generalized one-particle Parton Distributions \cite{Diehladd}.
These distributions can be  parametrized  as  \cite{Frankfurt}
\beq
G_1(x,Q^2,\Delta)=f_p(x,Q^2)F_{2g}(x,\Delta)
\eeq 
Here $f _p$ is  the conventional PDF  of the proton, while $F_{2g} $ is the so called two gluon formfactor \cite{Frankfurt}, which we shall parametrize in the 
dipole form
\beq
F_{2g}(\Delta)=\frac{1}{(1+\Delta^2/m_g^2)^2}
\eeq
Here $m_g$ is the parameter of order of the nucleon mass, i.e. 1 GeV,charactering the radius of gluon distribution in the proton.
The parameter $m_g$ depends on the transverse scale very weakly, and this dependence will be neglected,
 It has also  a weak dependence on $x_B$:
 \beq
 m_{g}^2=8/\delta ,
 \eeq
 where
 \beq
 \delta=max(0.28fm^2, 0.31fm^2+0.014fm^2\log(0.1/x)).
 \eeq
 This form was  was determined from the analysis of the exclusive $J/\Psi$ diffractive photoproduction at HERA \cite{Frankfurt}.

For the nucleon-nucleon collisions  the cross section of the DPS is given by 
\begin{eqnarray}
\sigma &=&f_p(x_1,Q_1^2)f_p(x_2,Q_2^2)f_p(x_3,Q_1^2)f_p(x_4,Q_2^2)\int \frac{d^2\Delta }{(2\pi)^2}\prod_{i=1,4}F_2g (\Delta,x_i)\frac{d\sigma_{24} }{dt_2} \nonumber\\[10pt]
&=&\frac{1}{28\pi}m^2_g f_p(x_1,Q_1^2)f_p(x_2,Q_2^2)f_p(x_3,Q_1^2)f_p(x_4,Q_2^2)\frac{d\sigma_{24} }{dt_2} .\nonumber\\[10pt]
\label{sigma1}
\end{eqnarray}
and the last equality in Eq. \ref{sigma1} is valid if we neglect the dependence of $m_g$ on x.
\par Consider now the case of the resolved photon-- proton collisions.
Due to the vector dominance we can expect that the GPD of the resolved photon is proportional to GPD of the vector meson.
We shall  assume the same factorization for the GPD$_1$ in the mean field approximation as for the proton.
\beq
G_!(x_1,\Delta,Q)_{\rm photon}= f(x,Q^2)_{\rm photon}F_{2g}^\gamma (\Delta,x)
\eeq
where in agreement with the vector dominance we put $F_{2g}^\gamma (\Delta)=F_{2g}^\rho (\Delta)$.

The transition formfactors for mesons decrease as $1/\Delta^2$ for  large transverse scale $\Delta$,
and the only relevant  parameter with dimension of mass is the  $\rho$ meson mass $m_\rho$.
Consequently we use a simple model for the two gluon formfactor of photon/vector meson:
\beq
F_{2g}^{\rho}(\Delta)=\frac{1}{(1+\Delta^2/m^2_\rho)}.
\eeq
This is confirmed by the  analysis of  the meson one particle Generalized Parton Distribution functions \cite{Ruiz}.
Similar formula was used in \cite{jimmy1}.
For \ss $\,\,$ processes we obtain, instead of factor $m^2_g/(28\pi)$ for DPS in pp, which is of order $0.14/(4\pi)$ GeV$^2$,
after the integration over $\Delta$, the factor 
\begin{eqnarray}
U&=&\int \frac{d^2\Delta}{(2\pi)^2}  F_{2g} (\Delta)^2\frac{1}{(1+\Delta^2/m^2_\rho)^2}\nonumber\\[10pt]
&=&\frac{m^2_gm^2_\rho}{4\pi}((m^2_g - m^2_\rho) (3m_g^6 - 5 m_g^2 m^4_\rho+ 13 m^4_g m_\rho^2 + m_\rho^6) + 
 12 m_g^4 m_\rho^2 \log(m^2_\rho/m^2_g))/(3 (m^2_g - m^2_\rho)^5)\nonumber\\[10pt]
 \end{eqnarray}
Here  we neglected  the dependence of  $m^2_g$  on $x_3,x_4$. which we take into account in numerical simulations.
In this approximation, taking the mass of $\rho $ meson 800 MeV we obtain $U\sim  0.17/(4\pi)$ GeV$^2$.
For \ss $\,\,$ contribution of resolved photon  we obtain finally
 \beq
d\sigma=n_\gamma(y)\alpha_{\rm em}\frac{f^2_\rho}{(4\pi)\kappa} f_\rho(x_1,Q_1)f_\rho(x_2,Q_2, )f_p(x_3,Q_1)f_p(x_4,Q_2)U(x_3,x_4)\frac{d\sigma_{13}}{dt_1}\frac{d\sigma_{24} }{dt_2}
\label{ff1}
\eeq
where 
\beq U(x_3,x_4)=\int \frac{d^2\Delta}{(2\pi)^2} F_{2g}(x_3,\Delta)F_{2g}(x_4,\Delta)\frac{1}{(1+\Delta^2/m_\rho^2)^2}\label{ff}
\eeq
In numerical calculations we shall use the PDFs from Refs. \cite{GR3},\cite{GRV} for photon and nucleon respectively. Also note that in DPS
we transform photon into $\rho$ meson only once, and then actually extract two partons from the wave function of $\rho$ meson,
so the pdfs in  Eq. \ref{ff1} are those of the vector meson,
\beq 
f_\gamma (x,Q^2)\sim \alpha_{\rm em} \frac{f^2_\rho}{(4\pi)\kappa} f_\rho(x,Q^2)
\eeq
Note  that in  Eq. \ref{ff1}  $\alpha_{em}$ appears in Eq. \ref{ff1} only once \cite{jimmy1}. The factor $f^2_\rho/(4\pi)\sim 0.5$ GeV$^2$ The coefficient $\kappa$ is between 1 and 2.
and takes into account the contribution of the  states other than $\rho$ meson into vector dominance., It is equal to 2 in the parametrization of \cite{GR3}
In the Leading Logarithmic Order .

\subsubsection{$1\otimes 2$ $\,\,$ mechanism.}
 \par Consider now   the $1\otimes 2$  $\,\,$ processes.
 Recall that these  processes will be now not symmetric between target and projectile, as in pp case,  as  it was mentioned above:,   one process will correspond to parton from the resolved photon splitting in 2 partons,
 we call them \12 processes,and 
 the second type of these  processes corresponds to parton in the nucleon PDF splitting in two.  For  the \12 processes we get the 
geometric  enhancement coefficient  the same as in  $1\otimes 2$ processes  in pp collisions: 
\beq
U(x_3,x_4)=\int \frac{d^2\Delta}{(2\pi)^2} F_{2g}(\Delta.x_3)F_{2g}(\Delta,x_4)=\frac{m^2_g}{3(4\pi)}, 
\eeq
where the last equality arises if we neglect dependence of two gluon formfactor on x.
The cross section for the \12 process
 \beq
 \frac{d\sigma^u}{dy}=n_\gamma (y)_1D^A_{BC}(x_1,x_2, Q_1^2,Q_2^2)U(x_3,x_4)f_p(x_3,Q_1^2)f_p(x_4,Q^2_2)\frac{d\sigma_{13}}{dt_1}\frac{d\sigma_{24} }{dt_2}
 \eeq
 
 where $f_p$ are PDFs of the proton. The. function  $_1D^A_{BC}(x_1,x_2,Q_1^2,Q_2^2)$ is the part of the photon  double GPD corresponding  to the split of the parton A in the resolved photon into two:
 partons B and C.
\begin{eqnarray}
_1D^A_{BC}(x_3,x_4,Q_1,Q_2)&=&\int \frac{d^2k_t}{(2\pi)^2}\Phi^A_{BC}(z)f_\gamma^A(y,k^2_t)\nonumber\\[10pt]
&\times&D_{AC}(x_1/(yz),k^2_t,Q_1^2)D_{AB}(x_2/(y(1-z)),k^2_t,Q_2^2)/(y^2z(1-z)k^2_t)\nonumber\\[10pt]
\label{tur2}
\end{eqnarray}

Here  $\Phi^A_{BC}$ is the splitting vertex, which is equal to the corresponding $A\rightarrow BC$ DGLAP kernel \cite{DDT}, and $ f_p$  is.  the  PDF of the splitting parton in the resolved photon.. 
The corresponding formfactor  U is the same as  for the  $1\otimes 2$  processes in pp collisions. The functions $D_{AB}$ and $D_{AC} $ are the fundamental solutions 
of the DGLAP equations \cite{DDT,BDFS2,BDFS3}. The integration  in $k_t$ in Eq. \ref{tur2} is carried from $Q_0^2\sim 0.5$ GeV$^2$, to the minimum of $Q_1^2,Q_2^2$,
\cite{BDFS2,BDFS3}. For the heavy quark final states we start the integration from the scale $\sim m_c=1.3$ GeV--the mass of the charm quark.

 \par  Consider now \21 processes. Their cross section is 
 \beq
 \frac{d\sigma^d}{dy}=n_\gamma (y)\alpha_{\rm em}\frac{f^2_\rho}{4\pi \kappa}f_\rho(x_1,Q^2_1)f_\rho (x_2,Q^2_2)) _1D(x_3,x_4,,Q_1^2,Q_2^2)U(x_1,x_2)\frac{d\sigma_{13}}{dt_1}\frac{d\sigma_{24} }{dt_2}
 \eeq
 where $f_\gamma$ are thenPDFs of the photon, and   $_1D(x_3,x_4,Q_1^2,Q_2^2)$ is the part of the parton GPD corresponding, to split of the parton iA n proton into two:
\begin{eqnarray}
_1D^A_{BC}(x_3,x_4,Q_1,Q_2)&=&\int \frac{d^2k_t}{(2\pi)^2}\Phi^A_{BC}(z)f_p^A(y,k^2_t)\nonumber\\[10pt]
&\times&D_{AC}(x_3/(yz),k^2_t,Q_1^2)D_{AB}(x_4(y(1-z)),k^2_t,Q_2^2)/(y^2z(1-z)k^2_t)\nonumber\\[10pt]
\end{eqnarray}
where $\Phi$ is the splitting vertex, and $f_p$  is the  PDF of the splitting parton in the  nucleon. The corresponding formfactor  U is now
just $m^2_\rho/(4\pi)$. 
 Note that if we assume constant $m_g\sim 1$ GeV, we can estimate the contribution of  \12 , \21 relative  \ss $\,\,$ processes for geometric enhancement coefficients..
 Indeed, comparing the formfactors we have the  enhancement of order f$\sim 1.5$ for \12 process relative to \ss $\,\,$ (in pp this enhancement was 7/3). On the other hand  for \21 processes we get the 
 enhancement factor 
 \beq
 \int \frac{d^2\Delta}{(2\pi)^2}\frac{1}{(1+\Delta^2/m^2_\rho)^2}=\frac{m^2_\rho}{4\pi}
 \eeq
 i.e. the geometric enhancement factor $0.64/0.17\sim 3.8$.  In the \12 $\,\,$ the corresponding ratio is $0.33/0.17\sim 2$.

\subsection{Direct Photon.}

We now consider the direct photon. The corresponding transition means,  see Fig. \ref{dps2} that the photon splits into a pair of quarks, with transverse momenta $k_t$ is  
 of order $2m_q$ or larger. The corresponding formalism was discussed 
 in \cite{BS,BlokSegev} .

The differential cross section for direct process for the case of photoproduction 
  and the final state of two hard processes each consisting of a gluon and quark jet is given by :

\begin{eqnarray}
 \frac{d\sigma}{dydx_1dx_2dx_3dx_4 dp_{1t}^2dp_{2t}^2}
&=&\frac{\alpha_{\rm em}^2N_c}{2}\int^{ 1-x_2}_{x_1}dz \int \frac{d^2k_t}{(2\pi)^2}\frac{y}{(k_t^2+m_c^2)^2}\nonumber\\[10pt]
&\times&(\vec k_{t}^2(z^2+(1-z)^2)(\frac{(1+(1-y)^2)}{y^2}\log(\frac{Q_{max}^2}{Q_{min}^2}))
-2m_e^2(\frac{1}{Q^2_{max}}-\frac{1}{Q^2_{min}})
\nonumber\\[10pt]
&\times&D_{qA}(x_1/z,k_t^2,Q_1^2)D_{\bar qB}(x_2/(1-z),k_t^2,Q_2^2)\frac{1}{z(1-z)}\nonumber\\[10pt]
&\times&\frac{M^2}{(x_1x_3\sqrt{x_1x_3})16\pi (yW^2)^{3/2}\sqrt{x_1x_3yW^2-4Q_1^2}}\nonumber\\[10pt]
&\times&\frac{M^2}{(x_2x_4\sqrt{x_2x_4})16\pi (yW^2)^{3/2}\sqrt{x_1x_3yW^2-4Q_2^2}}\nonumber\\[10pt]
&\times& U(x_3,x_4)f(x_3,Q_1^2)f_g(x_4,Q_2^2)\nonumber\\[10pt]
\label{qcd}
\end{eqnarray}
In this section we use the following DGLAP vertices:
\begin{eqnarray}
\Phi^g_{ gg}&=&2N_c(z/(1-z)+(1-z)/z+z(1-z))\nonumber\\[10pt]
\Phi^g_{ q\bar q}&=&\frac{1}{2}( z^2+(1-z)^2)\nonumber\\[10pt]
\Phi^q_{ qg}&=& c_F(1+z^2)/(1-z)\nonumber\\[10pt]
\end{eqnarray}
The matrix elements of the corresponding hard processes $\vert M \vert^2$ are given in \cite{Webber}..

\subsection{$\gamma $A collisions}

The general expressions for a  DPS cross section on a nuclear target is \cite{BSW}
\beq
\frac{d\sigma}{dx_1x_2dx_3dx_4dp_{1t}^2dp_{2t}^2}=D(x_1,x_2,Q_1^2,Q_2^2)f_p(x_3.Q_1^2)f_p(x_4,Q_2^2)\frac{d\sigma}{dt_1}\frac{d\sigma}{dt_2}\int \frac{d^2\Delta}{(2\pi)^2} F'_A(\Delta,-\Delta )
\eeq
 where D is the two parton GPD of the projectile nucleon or photon,and 
\beq
F'_A(\Delta,-\Delta )=F_A(\Delta,-\Delta )+AU(\Delta).
\label{tup}
\eeq
A is the total number of the nucleons in the nucleus.
The formfactor  $F_A(\Delta,-\Delta )$  is the  nucleus body form factor, while  the formfactor $U$  is defined by Eq. \ref{ff}.
The first term corresponds to the processes when two  partons  from a projectile photon interact with 
partons 
 from the different nucleons in the nucleus, see Fig. \ref{N1}right,Fig.\ref{N2} right.

 The  second term in Eq.\ref{tup} corresponds to the case when the two partons from the projectile   interact 
with the same nucleon, see Fig. \ref{N1},\ref{N2}left. The first term is expected to dominate for heavy nuclei as it scales as$A^{4/3}$ \cite{ST,BSW}.
 \par For the nuclear target we have
 \beq
 F_A(\Delta,-\Delta)=F^2(\Delta), F(\Delta)=\int d^2b \exp(i{\vec \Delta}\cdot  {\vec b})T(b),
 \eeq
 where
 \beq
 T(b)=\int dz \rho_A(b,z)dz
 \eeq
 is the nucleus profile function, b is the impact parameter.
 The nuclear form factor integral is expressed through the  profile function as
  \beq G(A)=\int \frac{d^2\Delta}{(2\pi)^2}F(\Delta,-\Delta)=\int T^2(b)d^2b=\pi\int T^2(b)db^2,
  \eeq
  where T(b) is calculated using the conventional mean field nuclear density \cite{Torbard,Vinas}
 \beq
\rho_A(b,z)=\frac{C(A)}{A}\frac{1}{1 + \exp{(\sqrt{b^2 + z^2} - 5.5\cdot A^{1/3})/(2.8)}}.
\eeq
The factor $C(A)$ is a normalization constant
\beq
\int d^2bdz \rho_A(b,z)=A.
\eeq
 Here the  distance scales  are given in  GeV$^{-1}$.
 Note that the nuclear form factor is independent of $x_3, x_4$.
 \par Then the only difference for direct photon  from the expression for $\gamma p$ collisions \ref{qcd} 
 is substitution (see Fig. \ref{N2})
 \beq
 U\rightarrow AU+G(A)
 \eeq
 \begin{center}
\begin{figurehere}
 \includegraphics[height=4.5cm,angle=0]{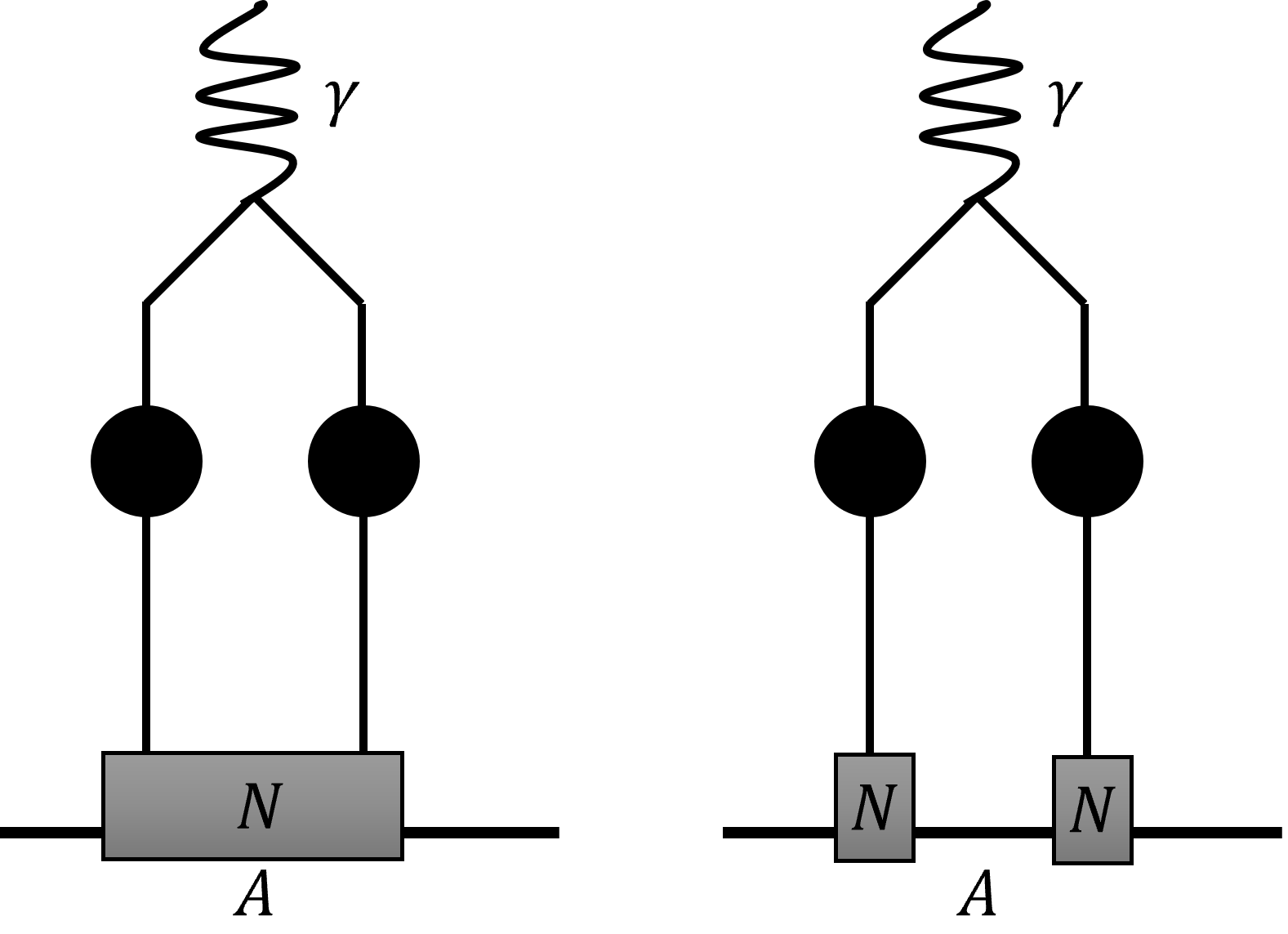}
\caption{\label{N1}The DPS in photon nuclei collisions: the direct photon}
\end{figurehere}
\end{center}

For the resolved  photon the term proportional G(A) includes only \ss contribution, since the \12 and \21 processes are rather small in this case
and can be neglected \cite{BSW},\cite{BS1}.
\begin{center}
\begin{figurehere}
 \includegraphics[height=4.5cm,angle=0]{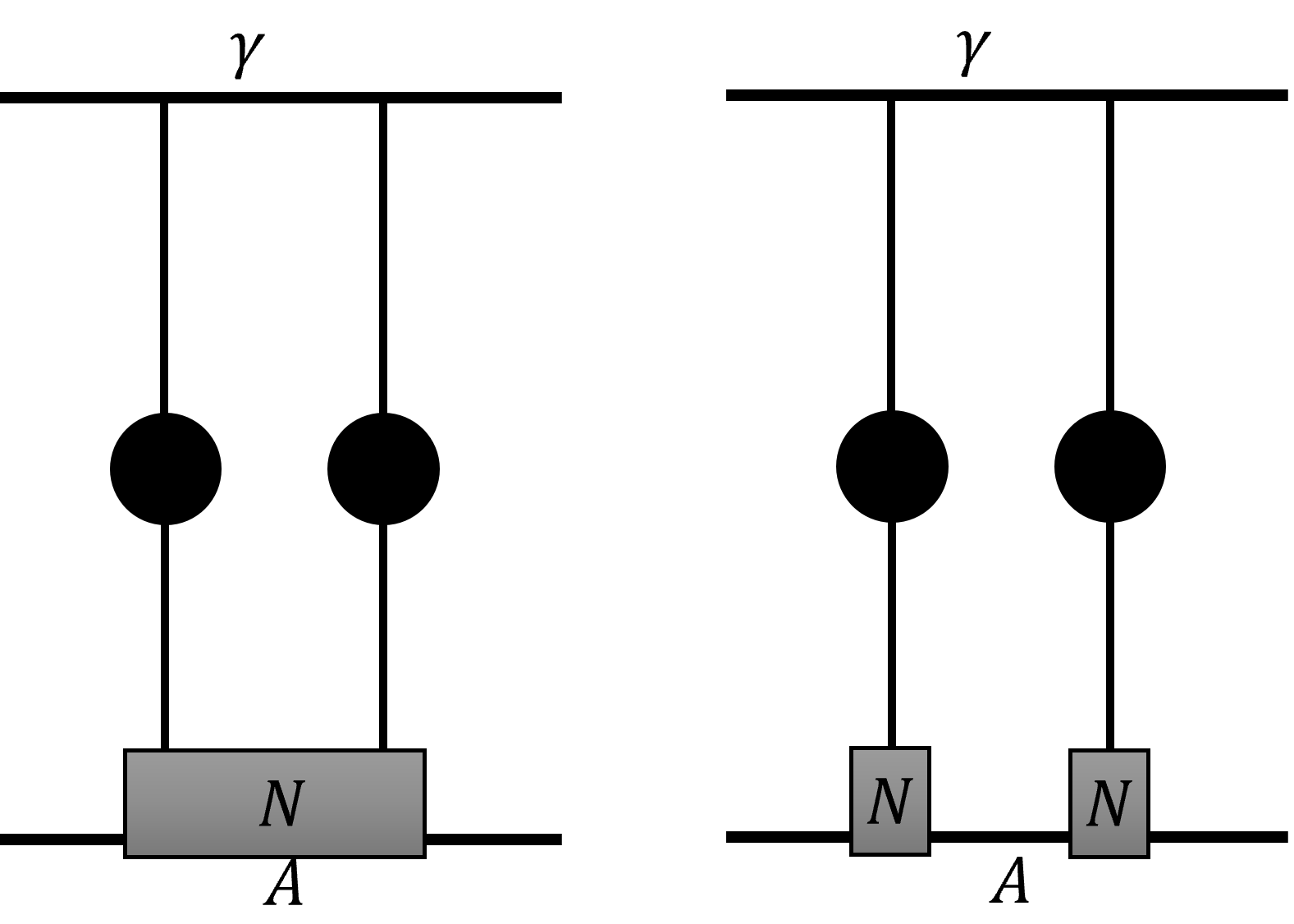}
\caption{\label{N2}The DPS in photon nuclei collisions: the resolved  photon}
\end{figurehere}
\end{center}
In the next chapters we carry our numerical calculations for $x_1=x_2=x_\gamma /2$. Nevertheless we expect that our results depend on $x_\gamma$
also for $x_1\ne x_2$, at least when they are not very far from each other.
					
\section{Numerics: Structure of the DPS with the Resolved Photon.}

We consider now resolved photons. We consider three different final states: two gluon minidijets and two gluon-two light and two heavy quark jets,
with $p_t\sim 3 $ GeV, considered in the previous paper.
We expect the qualitative features of these final states depend very weakly on  concrete $p_t$.
The general features are depicted in Figs. \ref{RG1},\ref{RG2}, \ref{RL1},\ref{RC1}.
We see that  in the $1\otimes 2$ processes even for small $p_t$  the increase the of DPS, by a factor of order 1.6, this enhancement becomes small only near the small x kinematic boundary
due to phase constraints. On the other side  it reaches maximum at $x_\gamma\sim 0.2-0.4$ and then decreases with the increase of $x_\gamma$.
\par Our results for resolved photon include conventional \ss process, that  are well known \cite{jimmy1, CR,jimmy2}, and  both \12 and \21  processes .
We depict our results  for two gluonic dijets final state in Fig. \ref{RG1}.
\begin{center}
\begin{figurehere}
 \includegraphics[height=5cm,angle=0]{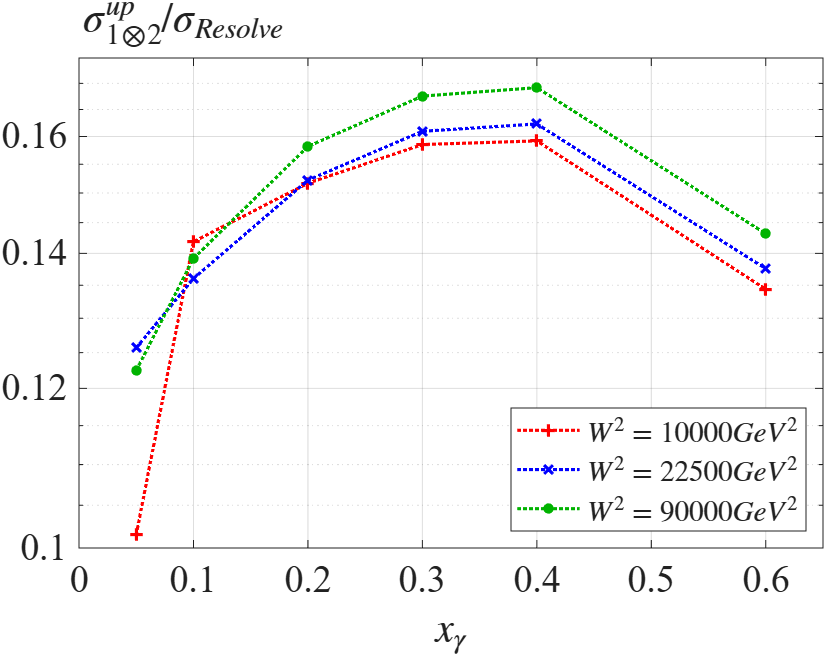}
  \includegraphics[height=5cm,angle=0]{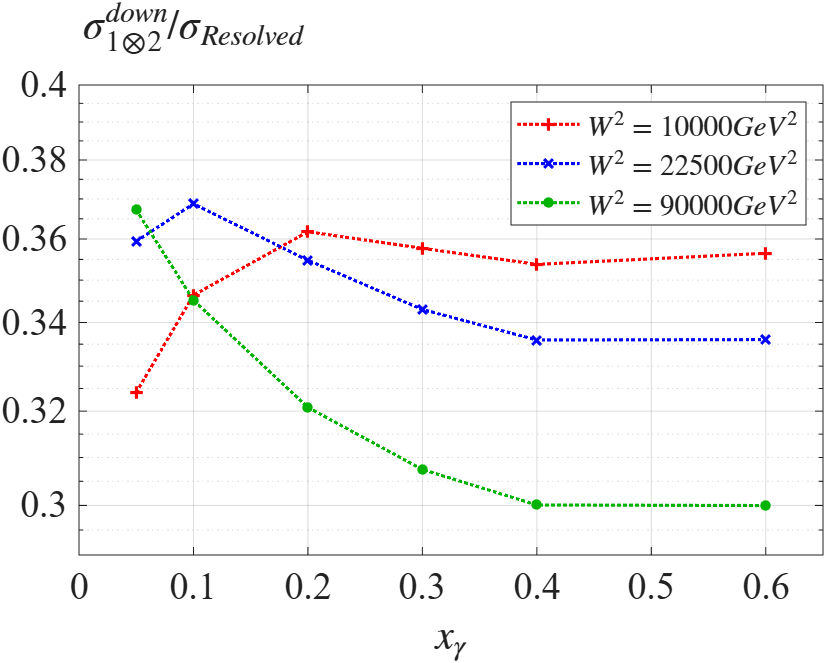}
  \includegraphics[height=5cm,angle=0]{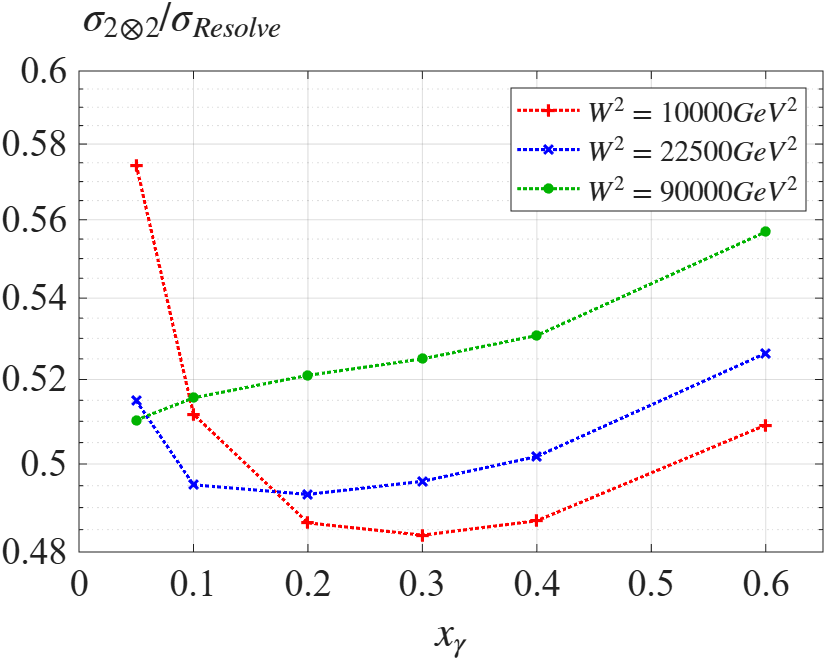}
\caption{\label{RG1} The ratio of \12 ,\21,\ss processes to total cross section of DPS for resolved photon and final state of  2 gluonic dijets.}
\end{figurehere}
\end{center}

We depict our results for 4 gluonic jets  density in Fig. \ref{RG2}. 
\begin{center}
\begin{figurehere}
  \includegraphics[height=5.5cm,angle=0]{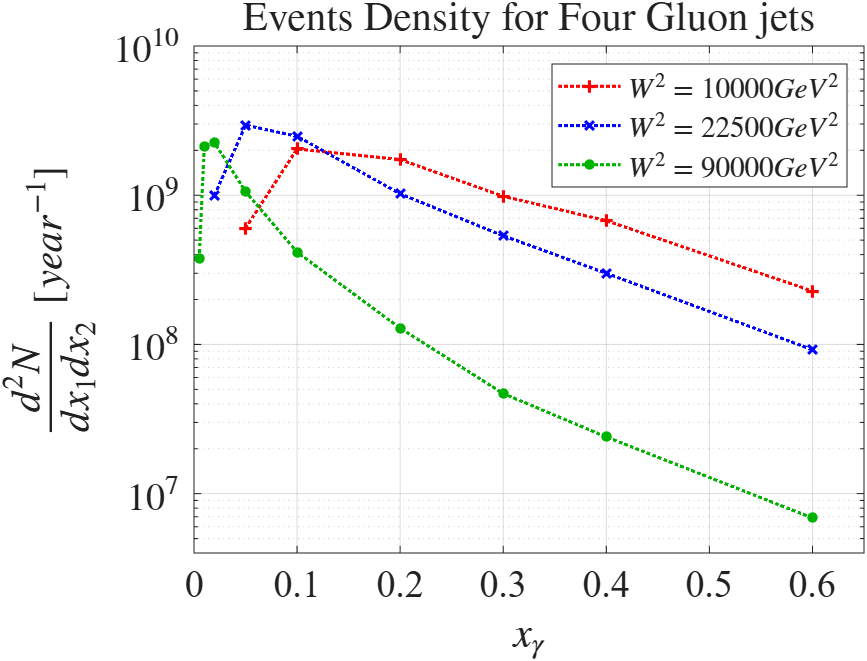}
\caption{\label{RG2}  The density of 4 gluonic minijets DPS}
\end{figurehere}
\end{center}
This density that characterisers a total number of DPS events has a maximum at $x_\gamma\sim 0.05-0.2$ depending on energy, and decreases for smaller and larger $x_\gamma$ due to phase constraints.

We depict our results for two gluonic and two light quark jets in Fig. \ref{RL1}
\begin{center}
\begin{figurehere}
 \includegraphics[height=5cm,angle=0]{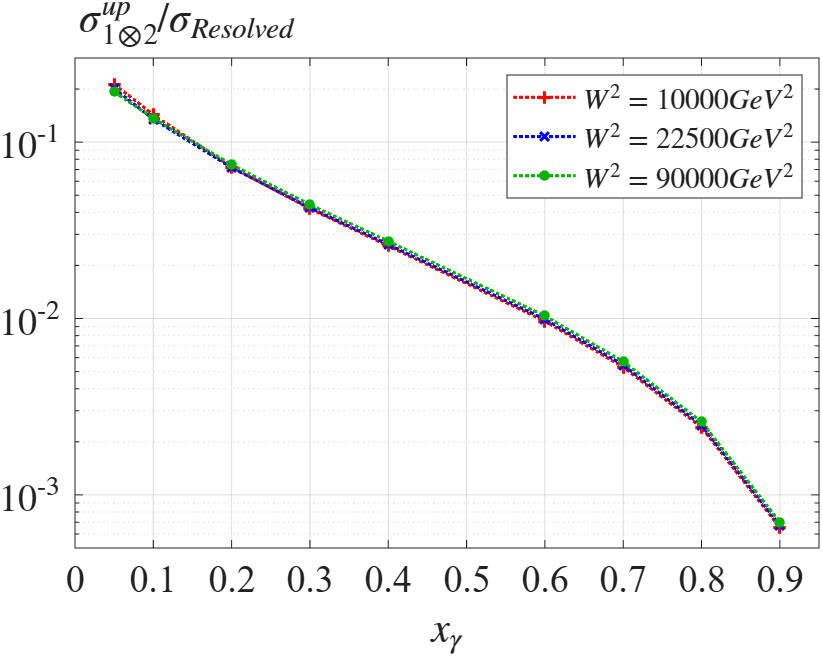}
  \includegraphics[height=5cm,angle=0]{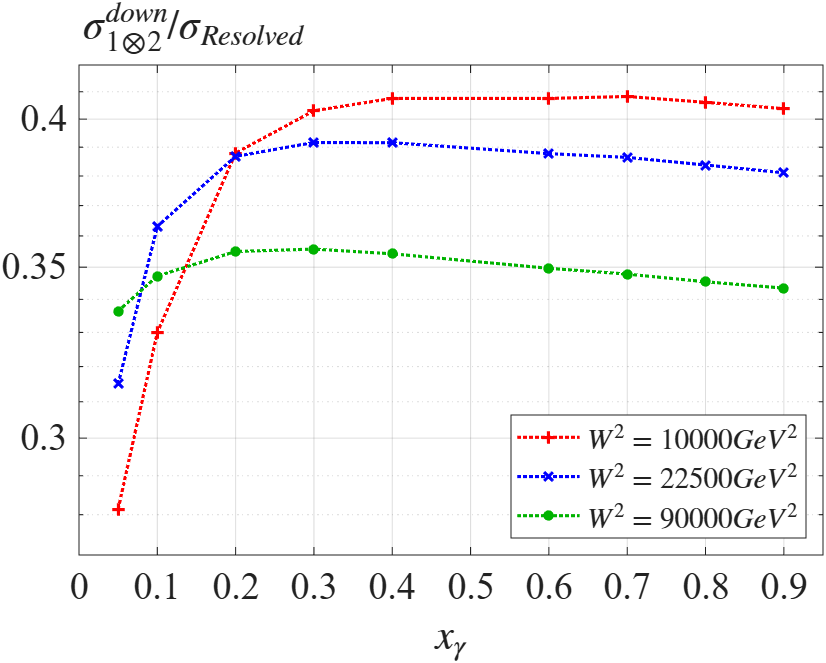}
  \includegraphics[height=5cm,angle=0]{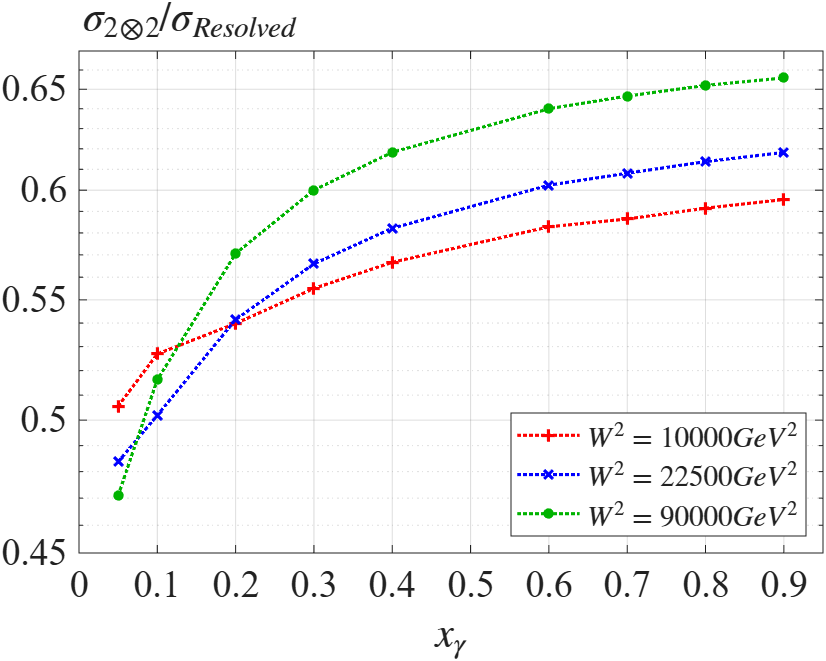}
\caption{\label{RL1} The ratio of \12 ,\12,\ss processes to total cross section of DPS for resolved photon and two hard processes, each of them including light quark and gluon jet.}
\end{figurehere}
\end{center}
We depict our results for two gluonic and two charmed quark jets in Fig. \ref{RC1}
\begin{center}
\begin{figurehere}
 \includegraphics[height=5cm,angle=0]{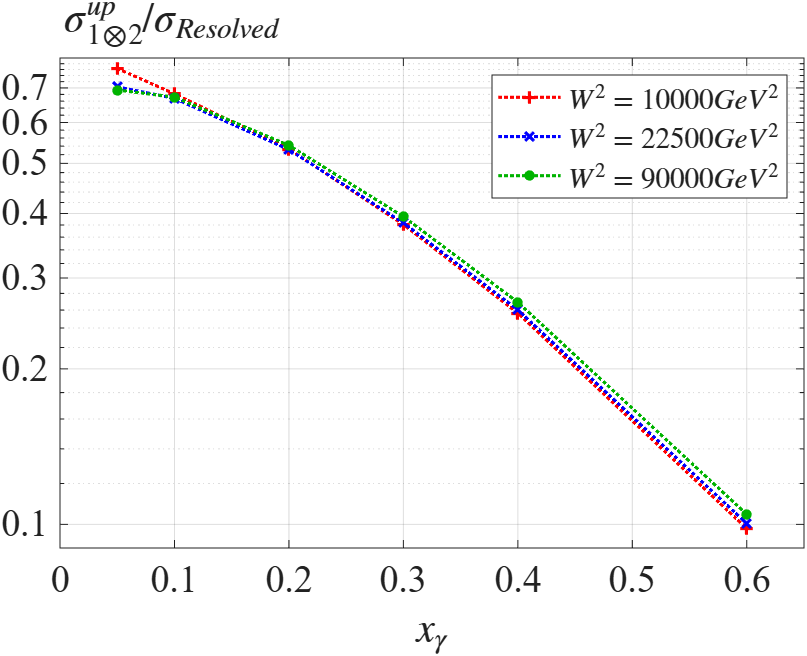}
  \includegraphics[height=5cm,angle=0]{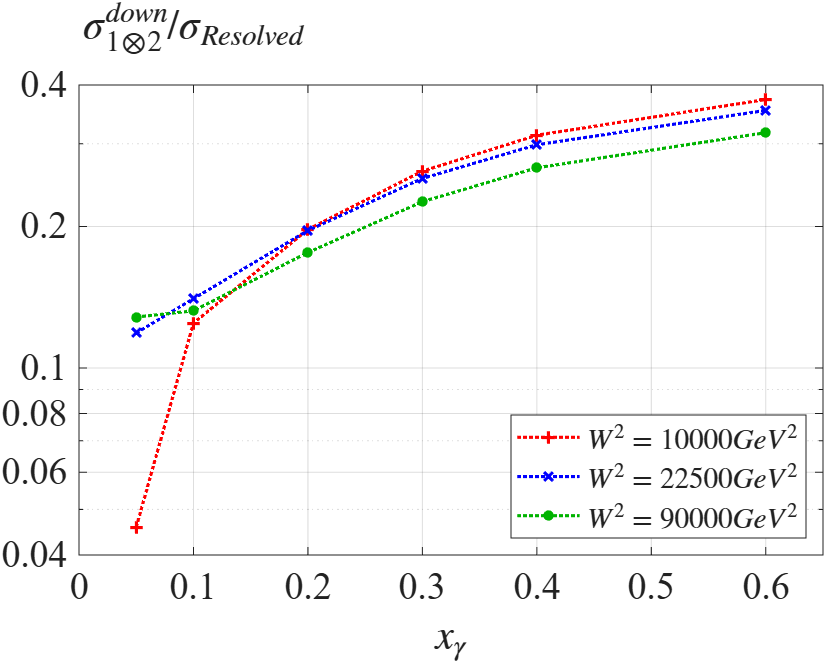}
  \includegraphics[height=5cm,angle=0]{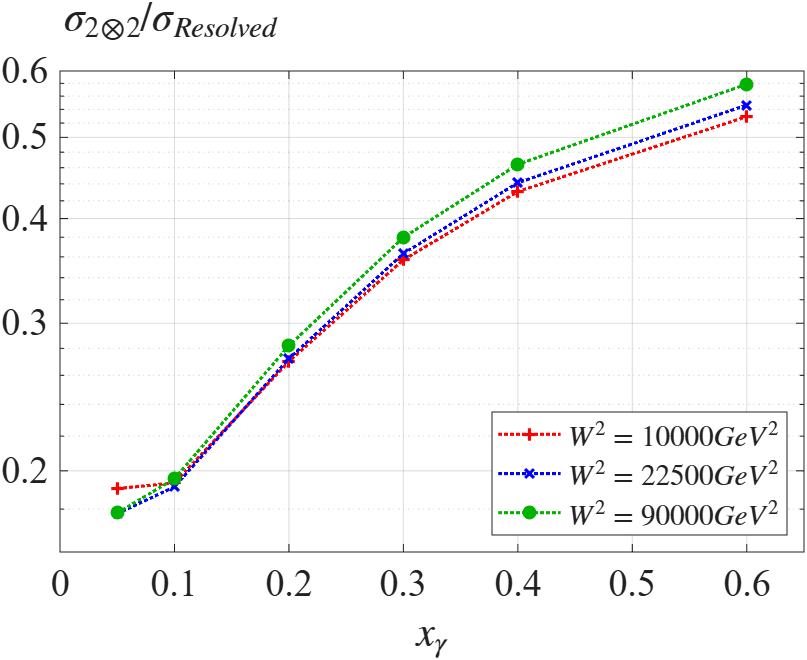}
\caption{\label{RC1} The ratio of \12 ,\12,\ss processes to total cross section of DPS for final state of two hard processes each with gluonic and charmed jets in the final state.}
\end{figurehere}
\end{center}
Note that all these results were obtained using vector dominance.
\par The results for light quark and charmed quark density similar to that in Fig. \ref{RG2} are depicted in the next section in comparison with the same quantity for direct transitions.

\section{Numerics:  Resolved versus Direct Photon}..
We can now compare tour results for resolved photon with the ones for direct photon for final states where each of the hard processes includes gluon and light quark or charmed jet..
For final state with charmed jets we shall see that the direct photons overcome resolved ones for $x_\gamma\sim 0.3-0.4$ and then quickly become dominant, so that for 
$x_\gamma\ge 0.5 $ the entire DPS cross section is dominated by direct photons, whose contribution was discussed in detail in \cite{BlokSegev}. For the final states with gluon and light quark jet 
the contributions of direct and resolved photon become of the same order for large $x_\gamma$.
\par Consider first the ratio of direct photon to resolved photon contribution for  light quark+gluon dijets. We depict this ratio in Fig. \ref{L1}
\begin{center}
\begin{figurehere}
 \includegraphics[height=5.5cm,angle=0]{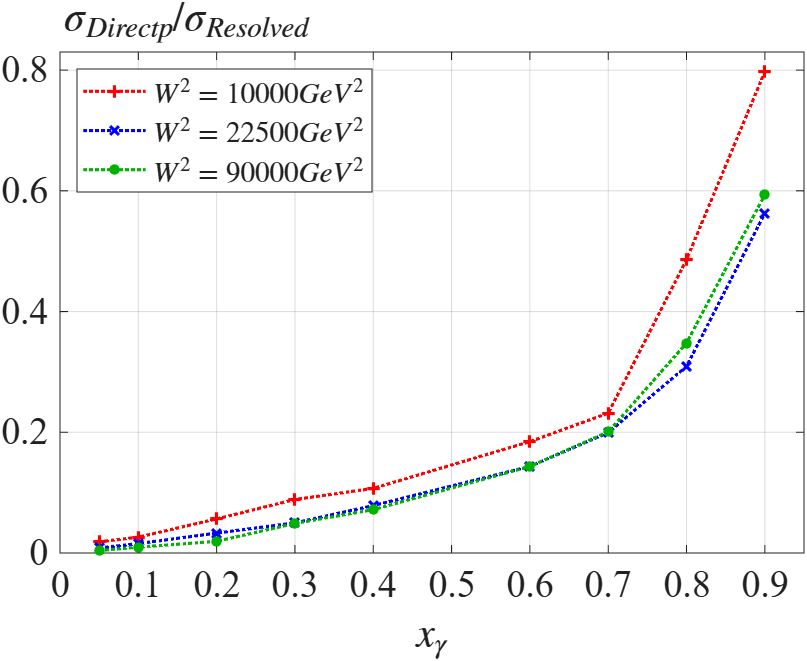}
  \caption{\label{L1} The ratio of direct to resolved contribution  for final state of gluon and light quark jet.}
\end{figurehere}
\end{center}
we also depict  in Fig. \ref{L2} relative densities of resolved and direct photon initiated processes at different energies
\begin{center}
\begin{figurehere}
 \includegraphics[height=5cm,angle=0]{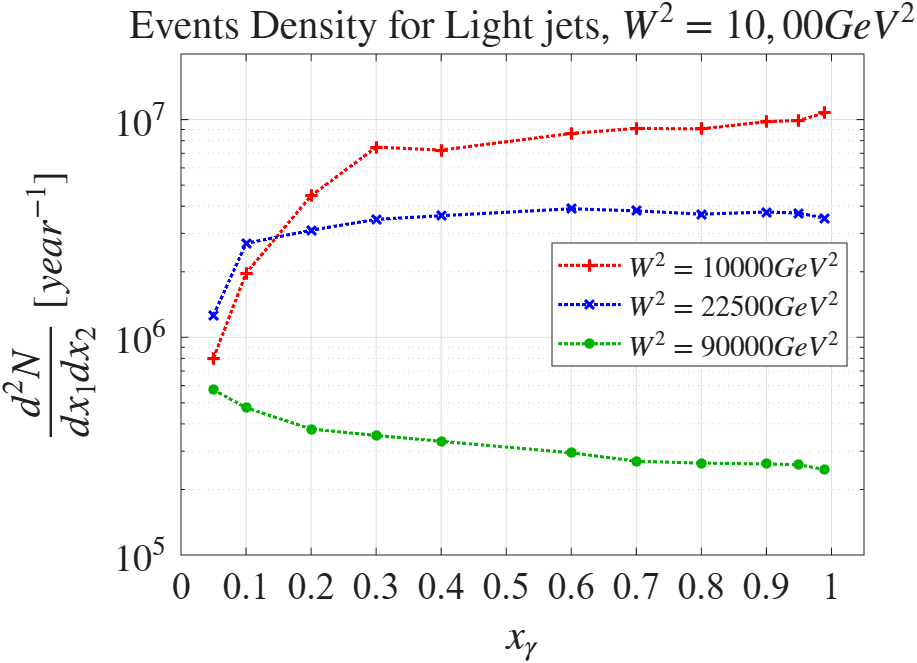}
  \includegraphics[height=5cm,angle=0]{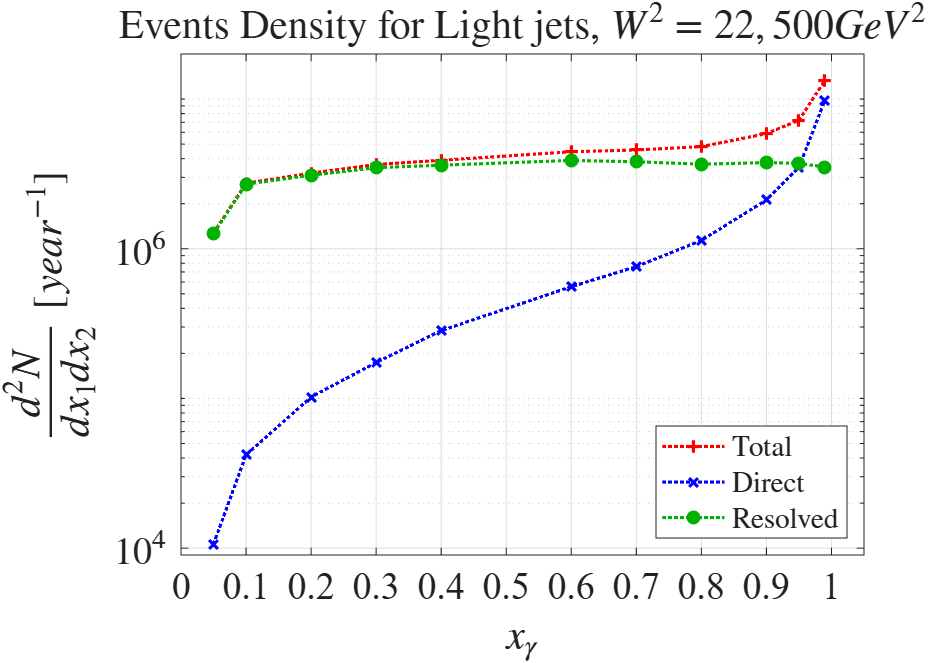}
  \includegraphics[height=5cm,angle=0]{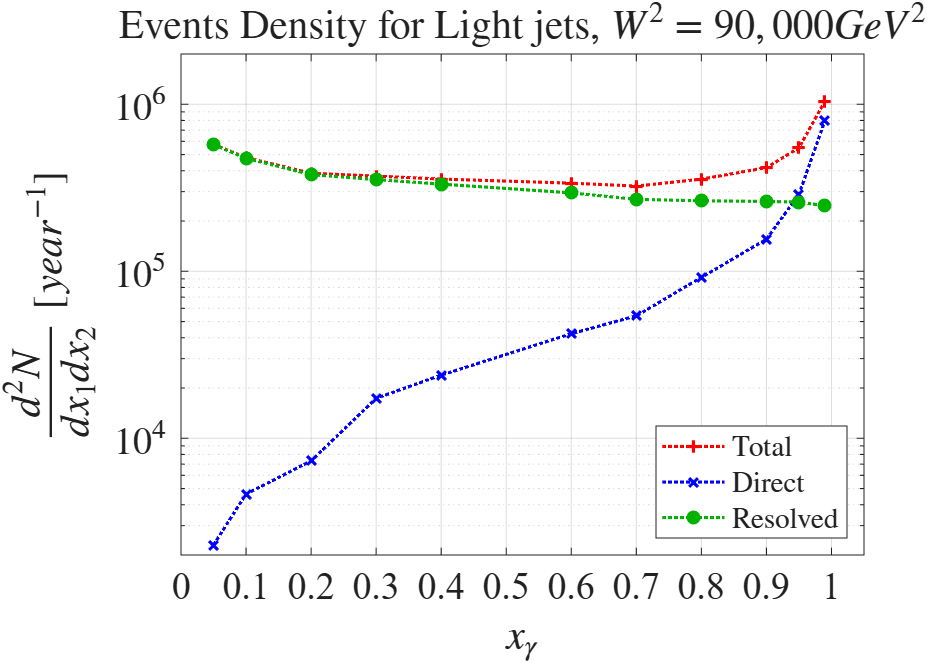}
  \caption{\label{L2} The  relative contributions of resolved and direct DPS cross sections for final state with the gluon  and  light quark jet.}
\end{figurehere}
\end{center}
and relation of direct to total  DPS  cross section for  these final states is depicted in Fig. \ref{L3}.
\begin{center}
\begin{figurehere}
 \includegraphics[height=5.5cm,angle=0]{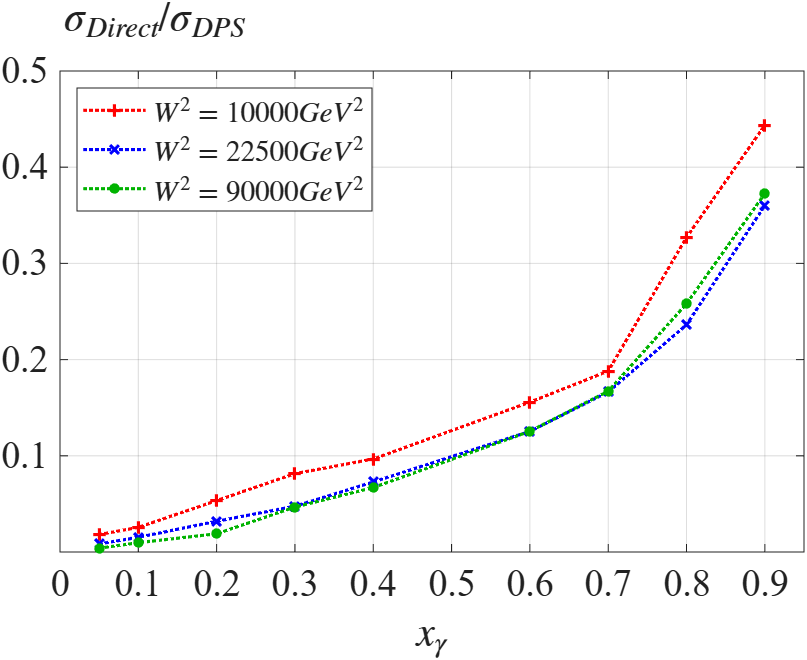}
  \caption{\label{L3} The ratio of direct to full DPS cross section  contribution  for final state with a gluon and light quark jet.s}
\end{figurehere}
\end{center}

For larger $x_\gamma$ the ratio tends to 1, although the numbers become very small due to phase constraints.

\par Consider  now final states with 2 hard processes, each of them consisting of gluonic and charmed jets.

\par Consider first the ratio of direct photon to resolved photon contribution for  this final state. This ratio is depicted in Fig \ref{C1}:
\begin{center}
\begin{figurehere}
 \includegraphics[height=5.5cm,angle=0]{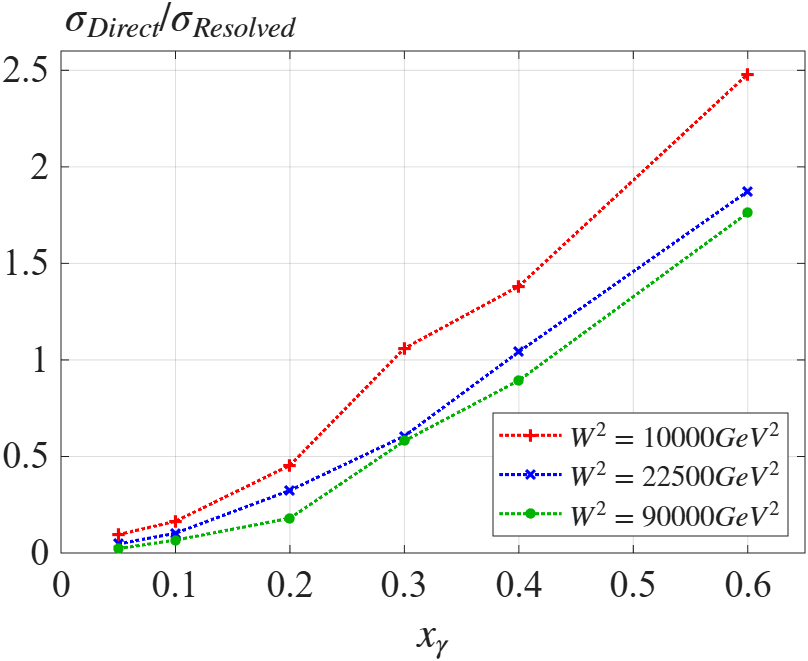}
  \caption{\label{C1} The ratio of direct to resolved contribution  for final state with gluon and charmed jets}
  \end{figurehere}
\end{center}
The  corresponding ratio to tal DPS cross section is depicted in Fig. \ref{C2}.
  
  \begin{center}
\begin{figurehere}
 \includegraphics[height=5.5cm,angle=0]{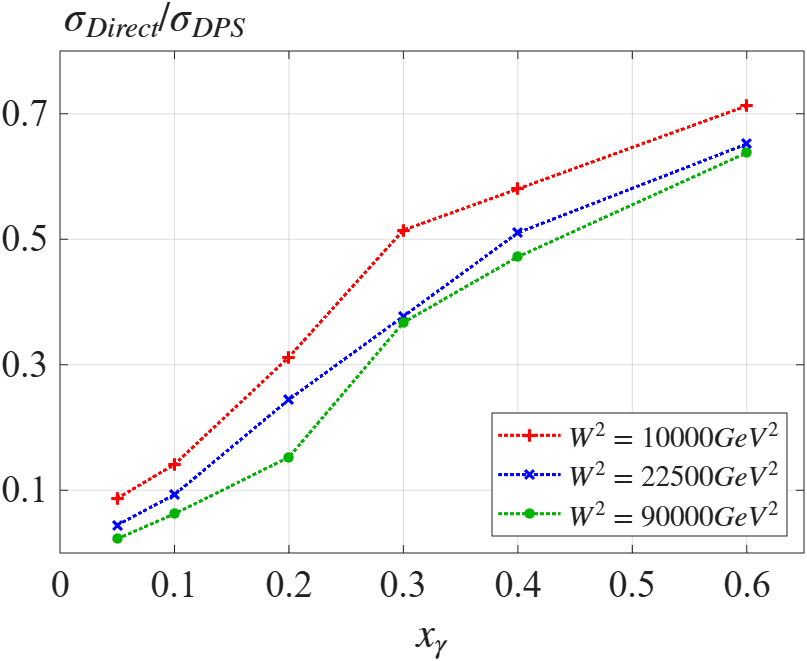}
\caption{\label{C2} The ratio of direct to total DPS  cross section for final state with gluon and charmed }
\end{figurehere}
\end{center}

we also depict relative densities of resolved, direct and total photon initiated DPS processes at different energies in Fig. \ref{C3}:
\begin{center}
\begin{figurehere}
 \includegraphics[height=5cm,angle=0]{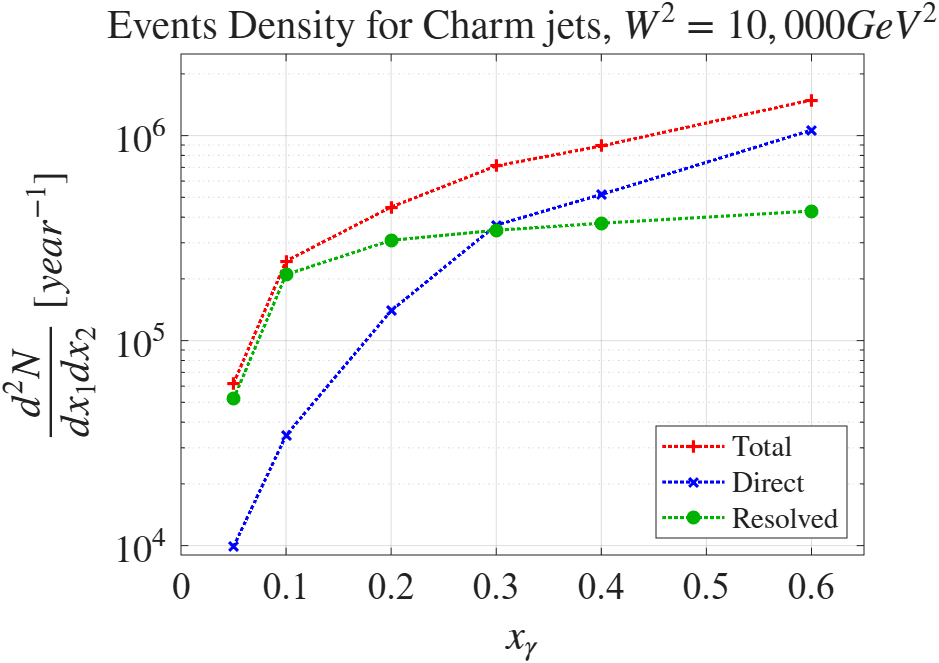}
  \includegraphics[height=5cm,angle=0]{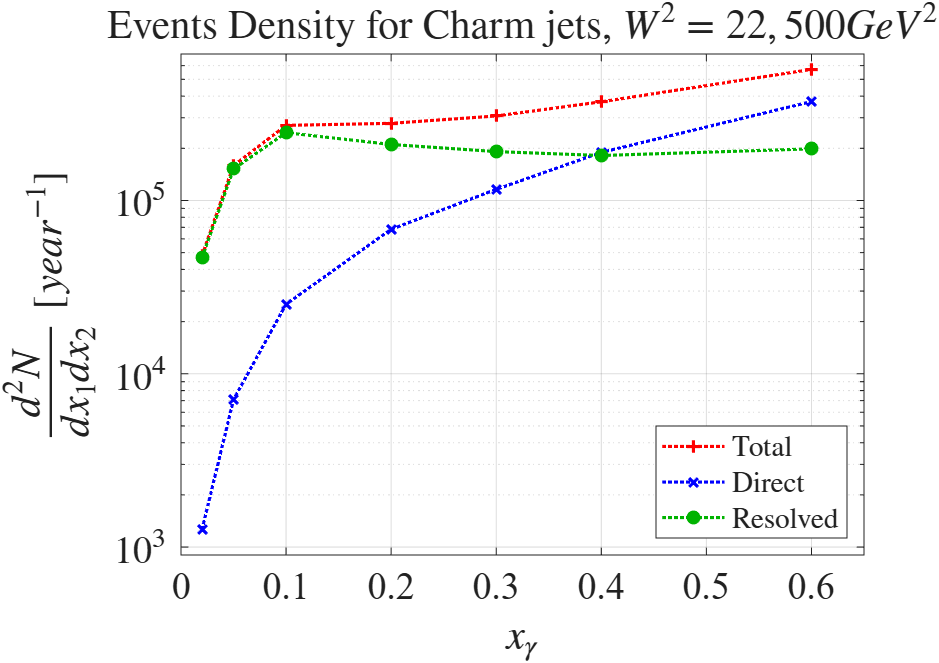}
  \includegraphics[height=5cm,angle=0]{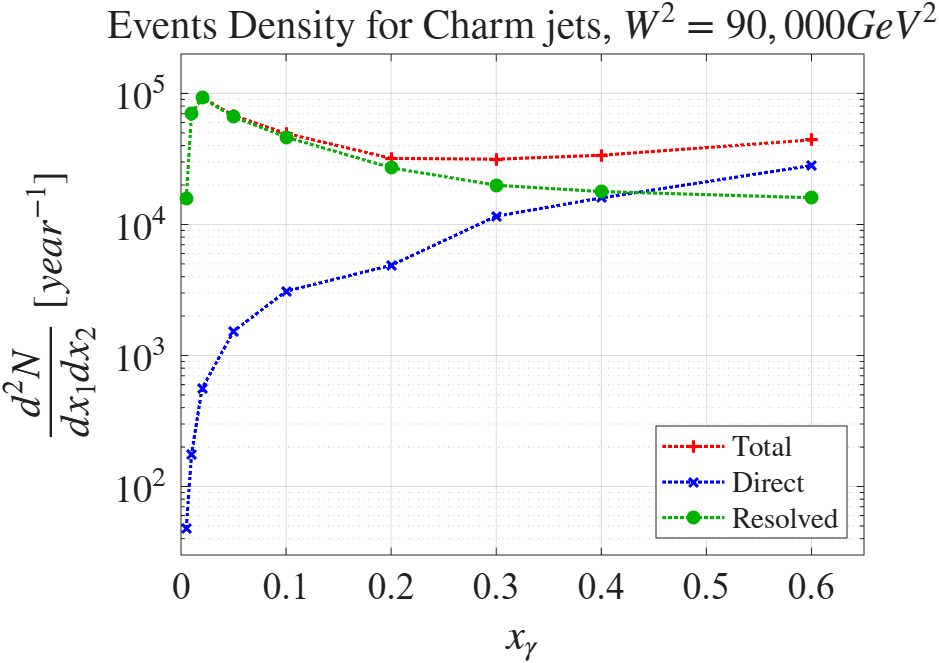}
  \caption{\label{C3} The  direct, resolved and total DPS as functions of energy and $x_\gamma$}
\end{figurehere}
\end{center}
 We see tthat at $x_\gamma\ge 0.3-0.4$ the direct photon contribution becomes dominant, Thus permitting the isolation of $1\otimes 2$ processes.

\section{DPS on Nuclei.}
\par In this case  there are two contributions: one is conventional enhanced by A , and another enhanced by $A^{4/3}$ which corresponds to photon interacting with two independent nuclei on the same impact parameter of his collision. 
The number of events on the nuclei as it was noted in the previous sections is 
\beq
\sim G(A)+AU(x_3,x_4)
\eeq
where the first term corresponds to interaction with two different nucleons. For Pb $G(A)\sim 12$.  As a result for the same luminosity and center of mass energy we have an increase 
of order 3 per nucleon in a number of events, as it was discussed in \cite{BlokSegev}.
\par The transition to nucleus slightly enhances the ratio $\sigma_{\rm direct}/\sigma_{\rm resolved}$, see Fig. \ref{N3}
The reason is that \12 mechanism is strongly suppressed and can be neglected, for the case of resolved photon interacting with two different nucleons \cite{BSW,BS1}.
The corresponding ratio $\sigma_{\rm direct}/\sigma_{\rm resolved}$ for the the case of final staes with two dijets-in one case they are charm and gluon, in the other case they are 
light quark and gluon, are depicted in Fig. \ref{N3}
\begin{center}
\begin{figurehere}
 \includegraphics[height=5.5cm,angle=0]{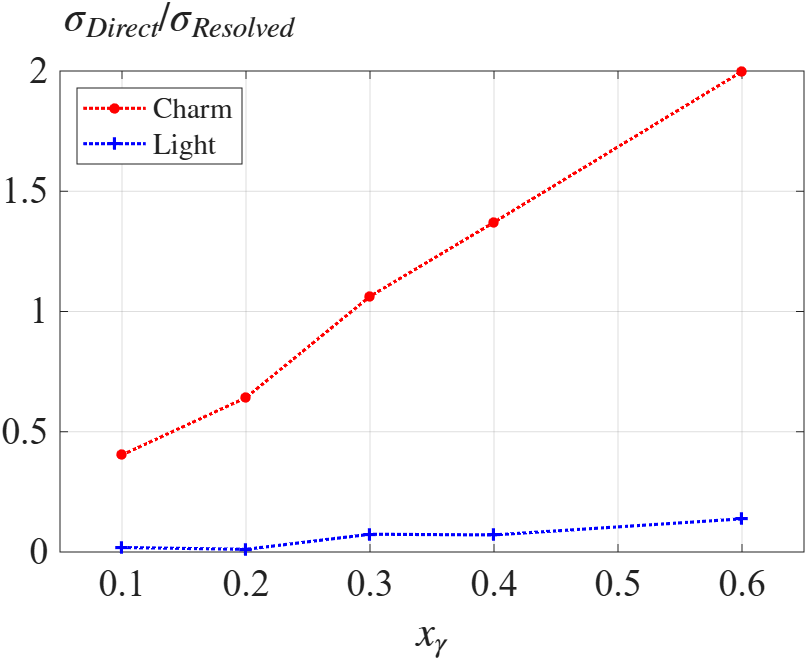}
  \caption{\label{N3} The  ratio of direct to resolved photon DPS  for photon--nucleus cross sections for final states with light and charm jets  as a function of  $x_\gamma$}
\end{figurehere}
\end{center}

\par On the other hand at EIC the characteristic center of mass energy for eA collisions is of 
order $5000 $ GeV$^2$, with luminosity $5*10^{33} \,\,cm^{-2}s^{-1}$. This leads to significant reduction 
in a number of events, as depicted in Fig. \ref{N4}, especially at small $x_\gamma\le 0.1$.
\begin{center}
\begin{figurehere}
 \includegraphics[height=5.5cm,angle=0]{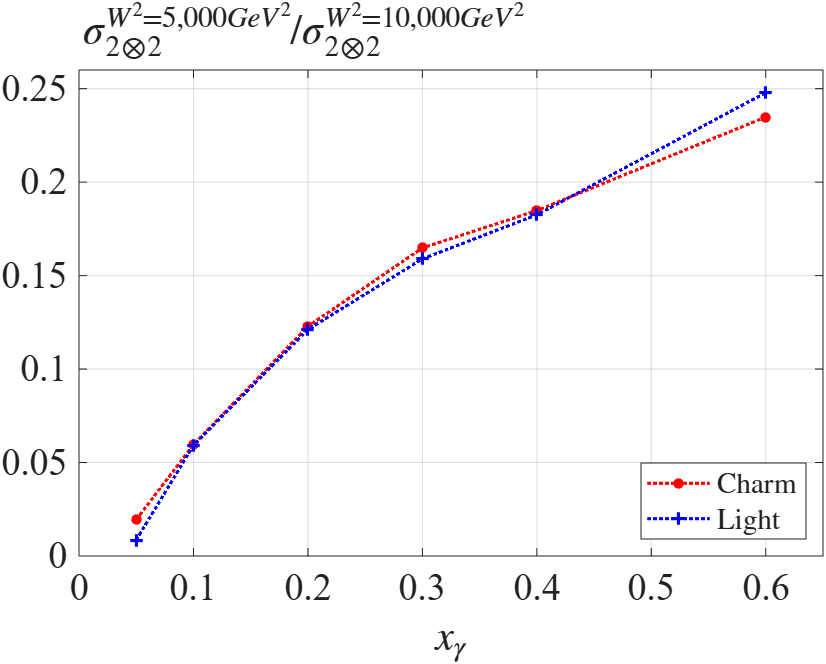}
   \caption{\label{N4} The  reduction of the phase space for photon nuclear collisions.}
\end{figurehere}
\end{center}

\section{Conclusion.}.
\par We have shown that
\par 1) the mean field contribution to the DPS initiated by resolved photon  is changed by a factor 1.6 for $x_\gamma\ge 0.1$ 
relative to  standard \ss $\,\,$mean field contribution.  The study of DPS in EIC will allow to reach kinematics of large $x$, in addition to 
central kinematics of order $x\sim 0.001-0.01$ that is available at LHC and TEVATRON.
\par 2) The two charm two gluon jets final state becomes dominant at $x_\gamma \sim  0.2-0.4$, depending on the energy,
\par 3) The direct photons become dominant even for smaller $x_\gamma$ for photoproduction on the nuclei for the same energy and luminosity.
\par Our results indicate that vector dominance does not give full description of the photon structure for large $x_\gamma$.


\begin{thebibliography}{99}
\bibitem{TreleaniPaver82}
N.\ Paver and D.\ Treleani,
  Nuovo Cim.\  A {\bf 70} (1982) 215.

\bibitem{mufti} M.\ Mekhfi, Phys. Rev. D{\bf 32}, 2371 (1985).

\bibitem{stirling} J.R.\ Gaunt and W.J.\ Stirling,
	  %``Double Parton Distributions Incorporating Perturbative QCD Evolution and
	  %Momentum and Quark Number Sum Rules,''
	  JHEP {\bf 1003}, 005 (2010)   %[arXiv:0910.4347 [hep-ph]]
	 
\bibitem{BDFS1}
  B.\ Blok, Yu.\ Dokshitzer, L.\ Frankfurt and M.\ Strikman,
  %``The Four jet production at LHC and Tevatron in QCD,''
  Phys.\ Rev.\  D {\bf 83}, 071501 (2011)
  %[arXiv:1009.2714 [hep-ph]].
  %%CITATION = PHRVA,D83,071501;%%
  
 \bibitem{Diehl} M.~Diehl,
	  %``Multiple interactions and generalized parton distributions,''
	  PoS D {\bf IS2010} (2010) 223
	  %[arXiv:1007.5477 [hep-ph]].
	  %%CITATION = POSCI,DIS2010,223;%%

\bibitem{stirling1} J.R.\ Gaunt and W.J.\ Stirling,
	  %``Double Parton Scattering Singularity in One-Loop Integrals,''
	  JHEP {\bf 1106},  048 (2011) %[arXiv:1103.1888 [hep-ph]].
	  
  \bibitem{BDFS2} B.\ Blok, Yu.\ Dokshitser, L.\ Frankfurt and M.\ Strikman,
  %``pQCD physics of multi-parton interactions,''
  Eur.\ Phys.\ J.\ C {\bf72}, 1963  (2012)
  %[arXiv:1106.5533 [hep-ph]].
  %%CITATION = ARXIV:1106.5533;%%
  
\bibitem{Diehl2} M.\ Diehl, D.\ Ostermeier and A.\ Schafer,
	  %``Elements of a theory for multiparton interactions in QCD,''
	  JHEP {\bf 1203} (2012) 089
	  %[arXiv:1111.0910 [hep-ph]].

\bibitem{BDFS3} B.\ Blok, Yu.\ Dokshitser, L.\ Frankfurt and M.\ Strikman,
  %``Origins of parton correlations in nucleon and multi-parton collisions','
 arXiv:1206.5594v1 [hep-ph] (unpublished).
 \bibitem{BDFS4}
 B.~Blok, Y.~Dokshitzer, L.~Frankfurt and M.~Strikman,
  %``Perturbative QCD correlations in multi-parton collisions,''
  Eur.\ Phys.\ J.\ C {\bf 74} (2014) 2926
  %[arXiv:1306.3763 [hep-ph]].


%\cite{Diehl:2017kgu}
\bibitem{Diehl:2017kgu}
  M.~Diehl, J.~R.~Gaunt and K.~Schönwald,
  %``Double hard scattering without double counting,''
  JHEP {\bf 1706} (2017) 083
  %doi:10.1007/JHEP06(2017)083
  %[arXiv:1702.06486 [hep-ph]].
  %%CITATION = doi:10.1007/JHEP06(2017)083;%%
  %27 citations counted in INSPIRE as of 08 Jan 2020

%\cite{Manohar:2012jr}
\bibitem{Manohar:2012jr}

  A.~V.~Manohar and W.~J.~Waalewijn,
  %``A QCD Analysis of Double Parton Scattering: Color Correlations, Interference Effects and Evolution,''
  Phys.\ Rev.\ D {\bf 85} (2012) 114009
  %doi:10.1103/PhysRevD.85.114009
  %[arXiv:1202.3794 [hep-ph]].
  %%CITATION = doi:10.1103/PhysRevD.85.114009;%%
  %99 citations counted in INSPIRE as of 08 Jan 2020  
  \bibitem{ST}M.~Strikman and D.~Treleani,
%``Measuring double parton distributions in nucleons at proton nucleus colliders,''
Phys. Rev. Lett. \textbf{88} (2002), 031801
doi:10.1103/PhysRevLett.88.031801
[arXiv:hep-ph/0111468 [hep-ph]].
%
\bibitem{BSW}
  B.~Blok, M.~Strikman and U.~A.~Wiedemann,
  %``Hard four-jet production in pA collisions,''
  Eur.\ Phys.\ J.\ C {\bf 73} (2013) no.6,  2433

\bibitem{book} Adv.\ Ser.\ Direct.\ High Energy Phys.\  {\bf 29} (2018) 2019, P. Bartalini and J. Gaunt Editors.
\bibitem{Yung}H. Yung, private communication.
\bibitem{jimmy1}J.~M.~Butterworth, J.~R.~Forshaw and M.~H.~Seymour,
%``Multiparton interactions in photoproduction at HERA,''
Z. Phys. C \textbf{72} (1996), 637-646
\bibitem{Yung1}H1 and Zeus Collaborations, Eur. Phys. J. C 72, 1995 (2012).
arXiv:1203.1170 [hep-ph]
\bibitem{Yung2} H. Jung (H1 and ZEUS Collaborations), in Proceedings 40th Inter-
national Symposium on Multiparticle Dynamics (ISMD 2010) 21– 25 Sept 2010, ed. by P. Van Mechelen,University of Antwerp, Belgium, pp 69–74. http://indico.cern.ch/conferenceDisplay.py? confId=68643. arXiv:1012.1554 [hep-ph]
\bibitem{UPC}
A.~J.~Baltz, G.~Baur, D.~d'Enterria, L.~Frankfurt, F.~Gelis, V.~Guzey, K.~Hencken, Y.~Kharlov, M.~Klasen and S.~R.~Klein, \textit{et al.}
%``The Physics of Ultraperipheral Collisions at the LHC,''
Phys. Rept. \textbf{458} (2008), 1-171
\bibitem{CR}
F.~A.~Ceccopieri and M.~Rinaldi,
%``Enlighting the transverse structure of the proton via double parton scattering in photon-induced interactions,''
Phys. Rev. D \textbf{105} (2022) no.1, L011501
doi:10.1103/PhysRevD.105.L011501
[arXiv:2103.13480 [hep-ph]].
\bibitem{jimmy2}
J.~M.~Butterworth, I.~Helenius, J.~J.~J.~Castella, B.~Pattengale, S.~Sanjrani and M.~Wing,
%``Modelling the underlying event in photon-initiated processes,''
SciPost Phys. \textbf{17} (2024) no.6, 158

\bibitem{BS} B.~Blok and M.~Strikman,
%``Double parton interactions in $\gamma p, \gamma A$ collisions in the direct photon kinematics,''
Eur. Phys. J. C \textbf{74} (2014) no.12, 3214
doi:10.1140/epjc/s10052-014-3214-7
[arXiv:1410.5064 [hep-ph]].
\bibitem{BlokSegev}B.~Blok and R.~Segev,
%``Double parton interactions initiated by direct photons in {\ensuremath{\gamma}}p and {\ensuremath{\gamma}}A collisions revisited,''
Phys. Rev. D \textbf{113} (2026) no.1, 014009
\bibitem{Feynman} R. Feynman, Photon-hadron interactions, Westview press, 1972.
\bibitem{Frixione:1993yw}
S.~Frixione, M.~L.~Mangano, P.~Nason and G.~Ridolfi,
%``Improving the Weizsacker-Williams approximation in electron - proton collisions,''
Phys. Lett. B \textbf{319} (1993), 339-345
\bibitem{Webber}R.~K.~Ellis, W.~J.~Stirling and B.~R.~Webber,
%``QCD and collider physics,''
Camb. Monogr. Part. Phys. Nucl. Phys. Cosmol. \textbf{8} (1996), 1-435
Cambridge University Press, 2011,
ISBN 978-0-511-82328-2, 978-0-521-54589-1
doi:10.1017/CBO9780511628788
\bibitem{Diehladd}M.~Diehl,
%``Generalized parton distributions,''
Phys. Rept. \textbf{388} (2003), 41-277
\bibitem{Frankfurt}
L.~Frankfurt, M.~Strikman and C.~Weiss,
%``Transverse nucleon structure and diagnostics of hard parton-parton processes at LHC,''
Phys. Rev. D \textbf{83} (2011), 054012
doi:10.1103/PhysRevD.83.054012
[arXiv:1009.2559 [hep-ph]].
\bibitem{Ruiz}W.~Broniowski and E.~Ruiz Arriola,
%``Impact parameter dependence of the generalized parton distribution of the pion in chiral quark models,''
Phys. Lett. B \textbf{574} (2003), 57-64.
\bibitem{GR3} M.~Gluck, E.~Reya and A.~Vogt,
%``Photonic parton distributions,''
Phys. Rev. D \textbf{46} (1992), 1973-1979
doi:10.1103/PhysRevD.46.1973
\bibitem{GRV}M.~Gl\"uck, E.~Reya and A.~Vogt,
%``Dynamical parton distributions revisited,''
Eur. Phys. J. C \textbf{5} (1998), 461-470
doi:10.1007/s100520050289
[arXiv:hep-ph/9806404 [hep-ph]].
\bibitem{DDT}Y.~L.~Dokshitzer, D.~Diakonov and S.~I.~Troian,
%``Hard Processes in Quantum Chromodynamics,''
Phys. Rept. \textbf{58} (1980), 269-395.

\bibitem{Torbard}C.~M.~Tarbert, D.~P.~Watts, D.~I.~Glazier, P.~Aguar, J.~Ahrens, J.~R.~M.~Annand, H.~J.~Arends, R.~Beck, V.~Bekrenev and B.~Boillat, \textit{et al.}
%``Neutron skin of $^{208}$Pb from Coherent Pion Photoproduction,''
Phys. Rev. Lett. \textbf{112} (2014) no.24, 242502
doi:10.1103/PhysRevLett.112.242502
[arXiv:1311.0168 [nucl-ex]].
\bibitem{Vinas}M.~Warda, X.~Vinas, X.~Roca-Maza and M.~Centelles,
%``Analysis of bulk and surface contributions in the neutron skin of nuclei,''
Phys. Rev. C \textbf{81} (2010), 054309
doi:10.1103/PhysRevC.81.054309
[arXiv:1003.5225 [nucl-th]].
\bibitem{BS1}B.~Blok and F.~A.~Ceccopieri,
%``$Z$ plus jets production via double parton scattering in $pA$ collisions at the LHC,''
Eur. Phys. J. C \textbf{80} (2020) no.8, 762.


\end{thebibliography}
\end{document}